\documentclass[useAMS,usenatbib]{mn2e}
\usepackage[pdftex]{graphicx}
\usepackage[pdftex,usenames,dvipsnames]{xcolor}
\usepackage{natbib}
\usepackage{aas_macros}
\usepackage{url}
\usepackage[none]{hyphenat}
\usepackage{times}
\usepackage[linktocpage=true,backref=true,pagebackref=false,bookmarks=true,bookmarksnumbered=False,colorlinks=true,linkcolor=NavyBlue,citecolor=NavyBlue,urlcolor=NavyBlue]{hyperref}
\usepackage[all]{hypcap}
\pdfpageattr {/Group << /S /Transparency /I true /CS /DeviceRGB>>}  

\title[Milliarcsec-scale radio emission of ULXs]{Milliarcsec-scale radio emission of ultraluminous X-ray sources: \\steady jet emission from an intermediate-mass black hole?}

\author[M. Mezcua et al.]
  {M.~Mezcua,$^{1,2,3}$\thanks{Email: mmezcua@iac.es.}
  S.A.~Farrell,$^{4}$
  J.C.~Gladstone,$^{5}$\thanks{Avadh Bhatia Fellow}
  A.P.~Lobanov,$^3$\thanks{Visiting Scientist, University of Hamburg / Deutsches Elektronen Synchrotron Forschungszentrum.}\\
  $^1$Instituto de Astrof\'isica de Canarias (IAC),
   		E-38200 La Laguna, Tenerife, Spain \\
  $^2$Universidad de La Laguna,
              Dept. Astrof\'isica, E-38206 La Laguna, Tenerife, Spain\\ 
  $^3$Max Planck Institute for Radio Astronomy,
              Auf dem H\"ugel 69, D-53121 Bonn, Germany \\
 $^4$Sydney Institute for Astronomy (SIfA), School of Physics, The University of Sydney, NSW 2006, Australia \\
 $^5$Department of Physics, University of Alberta, 11322-89 Avenue, Edmonton, Alberta, T6G 2G7, Canada }

\date{Released 2013 Xxxxx XX}

\pagerange{\pageref{firstpage}--\pageref{lastpage}} \pubyear{2013}

\def\LaTeX{L\kern-.36em\raise.3ex\hbox{a}\kern-.15em
    T\kern-.1667em\lower.7ex\hbox{E}\kern-.125emX}

\begin{document}

\label{firstpage}

\maketitle

\begin{abstract}
The origin of the high X-ray luminosities of most ultraluminous X-ray sources (ULXs) still remains poorly
understood. Most of the scenarios proposed to explain their nature assume that ULXs are powered by accretion on to a black hole
(BH). In this case, the detection of compact radio emission and the location of the ULXs in the Fundamental Plane
(X-ray versus radio-luminosity plane) can provide an estimate of the ULX BH mass and help address the question of their physical
nature. 
We present the results of a high-resolution (very long baseline interferometry) radio observational campaign aimed at detecting and studying compact radio emission in four ULXs with known radio counterparts. We find that one of the targets (N4559-X4) was previously misclassified: its low X-ray luminosity indicates that the source is not a ULX.
No milliarcsec-scale radio emission is detected for N4559-X4 nor for the ULXs N4490-X1 and N5194-X2, for which upper limits on the radio luminosities are estimated. These limits argue strongly against the presence of supermassive BHs in these three systems. For N4559-X4, the low X-ray luminosity and the ratio of the radio and X-ray luminosities suggest the presence of an X-ray binary.
Compact radio emission is detected for the ULX N5457-X9 within its \textit{Chandra} positional error, making N5457-X9 a potential intermediate-mass BH with steady jet emission. 
 \end{abstract}


\section{Introduction}
In the late 1970s, the \textit{Einstein} observatory detected a new class of extragalactic X-ray sources that were not spatially coincident with the nuclei of their host galaxies but had luminosities exceeding the Eddington luminosity of stellar-mass black holes (BHs; \citealt{1989ARA&A..27...87F}). Extensive studies of these ultraluminous X-ray sources (ULXs) have been made over the
past 30 years, but the physical mechanism triggering the high
luminosities observed is still under debate.  One of the
most appealing scenarios suggests that ULXs are intermediate-mass
black holes (IMBHs), with masses between 100 and $10^{5} M_\odot$
accreting at sub-Eddington rates
(e.g., \citealt{1999ApJ...519...89C}). These IMBHs could be the long sought after
seed BHs required to explain the existence of supermassive black holes
(SMBHs) in the early Universe (e.g., \citealt{2005ApJ...633..624V};
\citealt{2008MNRAS.383.1079V}). BHs with masses up to $\sim$100
$M_\odot$ could be formed from the death of very young and massive
(Population III) stars (\citealt{2010ApJ...714.1217B}) or from the core collapse of massive stellar clusters (\citealt{2004Natur.428..724P}), while BHs up to 10$^{6}$ $M_\odot$ could result from the direct collapse of
pre-galactic gas discs (\citealt{2006MNRAS.371.1813L}; see review by \citealt{2012Sci...337..544V}). According to BH mass scaling relationships (e.g., \citealt{2013ApJ...764..184M}), low-mass dwarf galaxies are expected to host nuclear IMBHs. Therefore, tidal stripping of merging satellite galaxies should result in the presence of wandering BHs [either IMBHs or off-nuclear active galactic nuclei (AGN)] in the haloes of large galaxies (\citealt{2008ApJ...687L..57V}; \citealt{2010ApJ...721L.148B}). ULXs could thus represent the nuclei of captured satellite galaxies (e.g., \citealt{2008MmSAI..79.1306L}). 

In another scenario, ULXs could be stellar-mass BHs that appear to radiate
at super-Eddington luminosities. Such luminosities could be produced
by pressure-dominated accretion discs with photon-bubble instabilities
(\citealt{2002ApJ...568L..97B}) or by beaming effects from either strong outflows
(geometric beaming; \citealt{2001ApJ...552L.109K}; \citealt{2007MNRAS.377.1187P}; 
\citealt{2008MNRAS.385L.113K,2009MNRAS.393L..41K}) or the axis of a relativistic jet 
pointing towards the observer (relativistic beaming;
\citealt{2002A&A...382L..13K}). The peculiar features of some ULX spectra 
(i.e., a soft excess and a break at high
energies) have been interpreted as support for the super-Eddington scenario (\citealt{2009MNRAS.397.1836G}).
X-ray timing analysis also seems to show that ULXs exhibit different properties from X-ray binaries in sub-Eddington accretions states, with apparent suppressed variability being observed (\citealt{2009MNRAS.397.1061H}). This led to the suggestion of a possible new `ultraluminous' state (\citealt{2009MNRAS.397.1836G}; \citealt{2010AIPC.1248..123R}; also see \citealt{2011NewAR..55..166F} for an extended review).
Examples of such super-Eddington accretors and `microquasars' can be found in
our Galaxy (e.g., SS433, \citealt{2006MNRAS.370..399B}; V4641,
\citealt{2002A&A...391.1013R}; GRS 1915+105,
\citealt{1994Natur.371...46M}), supporting the non-isotropic emission
scenario for ULXs with X-ray luminosities up to $10^{41}$ erg s$^{-1}$
(\citealt{2005MNRAS.356..401R}). Both the beaming and super-Eddington accretion
scenarios are in agreement with ULXs being commonly found in
star-forming galaxies, and are able to produce the extended
photoionized nebulae observed around some ULXs
(\citealt{2002MNRAS.332..764W}; \citealt{2003RMxAC..15..197P}; \citealt{2004MNRAS.351L..83K}; \citealt{2007ApJ...666...79L}; \citealt{2008AIPC.1010..303P};
\citealt{2009ApJ...697..950K}; \citealt{2011IAUS..275..325C};
\citealt{2011ApJ...731L..32M}; \citealt{2011AN....332..371R}). 
However, the non-isotropic radiation
scenario struggles to explain ULXs with X-ray luminosities above
$10^{41}$ erg s$^{-1}$. These extreme ULXs, which include the hyperluminous X-ray sources (HLXs) with $L_\mathrm{X} > 10^{41}$ erg s$^{-1}$, are the most probable sources for hosting IMBHs. 
So far, the strongest evidence for the presence of an IMBH was found in the ULX ESO 243-49 HLX-1 (\citealt{2009Natur.460...73F}). Another strong IMBH candidate (N4088-X1) was identified through very long baseline interferometry (VLBI) radio observations that detected compact radio emission (\citealt{2011AN....332..379M}), but follow-up high-resolution X-ray observations with \textit{Chandra} revealed that the radio and X-ray emission are not consistent (Mezcua et al. in preparation). 

Determining, or at least constraining, the ULX BH mass
($M_\mathrm{BH}$) is pivotal to reveal the physical mechanism
producing their high X-ray luminosities. Several methods can be used
to derive the $M_\mathrm{BH}$, the most reliable one being the optical
study of the stellar companion in high-mass X-ray binaries (for those ULXs with a high-mass companion), either using photometric (e.g., \citealt{2003RMxAC..15..197P}; \citealt{2010MNRAS.403L..69P}) or
spectroscopic techniques (e.g., \citealt{2006IAUS..230..293P};
\citealt{2009ApJ...697..950K};
\citealt{2011ApJ...728L...5C}; \citealt{2011AN....332..398R}). However, only
in a very few cases (the most nearby) have ULX optical counterparts been detected (e.g., \citealt{2011ApJ...737...81T}; \citealt{2013ApJS..206...14G}). Alternatively, some indirect methods can be used in the X-ray domain including: X-ray spectral fitting together with the
luminosity--temperature diagram (e.g., \citealt{2003Sci...299..365K}; \citealt{2003ApJ...585L..37M};
\citealt{2006MNRAS.371..673G}; \citealt{2009MNRAS.397..124G};
\citealt{2009ApJ...692..443S}; \citealt{2011ApJ...734..111D}; \citealt{2011ApJ...743....6S}; \citealt{2012ApJ...752...34G}), quasi-periodic oscillation
frequency scaling with $M_\mathrm{BH}$ (e.g.,
\citealt{2003ApJ...586L..61S}; \citealt{2007ApJ...660..580S}), the
correlation between X-ray photon index and $L/L_\mathrm{Edd}$
(\citealt{2008ApJ...682...81S}), or X-ray variability
(\citealt{2004A&A...423..955S}; \citealt{2009MNRAS.397.1061H}).

\begin{table*}
\caption{ULX target sources}
\label{table1}
\begin{minipage}{\textwidth}
\begin{center}
\begin{tabular}{cccccccccc}
\hline
\hline 
NGC      & ULX name    & 	RA (J2000)	& 	Dec. (J2000)  		&  	Most recent \textit{Chandra} 	&    Peak \textit{Chandra}	  &     Radio (VLA)   & Integrated     & VLA radio    & References  \\
	     &                     & 	 \textit{Chandra}& 	 \textit{Chandra}	& 	0.3--8 keV luminosity			&    0.3--8 keV luminosity   &   vs. X-ray offset     & VLA flux  	& emission      &               \\
	     &                     & 	        	         & 	         		       &   	(erg s$^{-1}$)				&    (erg s$^{-1}$)		  &      (arcsec)       & (mJy)       &              	&          \\
\hline
4490   	  & X1      	& 12 30 29.55  &  +41 39 27.6       &        9.80$\times10^{38}$		&        9.67$\times10^{39}$		&    1.1   		      &  7.17     &  Extended		&  1, 2, 3, 4	 \\
4559		  & X4$^{a}$	& 12 35 56.38 &  +27 59 25.8        &   	1.78$\times10^{38}$		&        0.21$\times10^{39}$		&    1.2      	      &  0.90     &  Compact		&  5, 6, 7, 9	   \\
5194        	  & X2 		& 13 29 50.67  & +47 11 55.2        &		5.58$\times10^{39}$		&        5.58$\times10^{39}$		&    0.4   		      &  3.20     &  Unresolved		&  2, 3, 4  \\
5457		  & X9$^{b}$ 	&  14 03 41.27 & +54 19 04.1        &		1.23$\times10^{38}$		&        2.92$\times10^{39}$		&    1.7   		      & 11.45    &  Extended		&  5, 6, 7, 8	   \\
\hline
\end{tabular}
\end{center}
{\bf Notes:}~$^{a}$ Name adopted from \cite{2006A&A...452..739S} and \cite{2005ApJS..157...59L}. The source was misidentified as a ULX (\citealt{2006A&A...452..739S}). It corresponds to source 8 in \cite{1997A&A...318..768V} and NGC4559-X7 in \cite{2011ApJS..192...10L}; $^{b}$ Name adopted from the X-ray catalog of \cite{2005ApJS..157...59L}. It corresponds to source XMM-13 in \cite{2004MNRAS.349..404J} and NGC5457-X50 in \cite{2011ApJS..192...10L}. References: (1) -- \cite{2004PThPS.155...27M}; (2) -- \cite{2004ApJS..154..519S}; (3) -- \cite{2011AN....332..384P}; (4) -- \cite{2011ApJ...741...49S}; (5) -- \cite{2006A&A...452..739S}; (6) -- \cite{2011ApJS..192...10L}; (7) -- \cite{2005ApJS..157...59L}; (8) -- \cite{2004MNRAS.349..404J}; (9) -- \cite{1997A&A...318..768V}.
\end{minipage}
\end{table*}

Detection and studies of radio counterparts to ULXs are an
excellent way to reveal their physical nature (where the radio emission is linked to relativistic jets from accreting BHs or shell-like structures from SNRs), by either directly
resolving them or providing a measure of their brightness temperature
and spectral properties. In particular, VLBI
radio observations have been shown to provide the best estimates of the ULX compact radio flux and brightness temperature at milliarcsec scales (e.g., \citealt{2011AN....332..379M}; \citealt{2013Natur.493..187M}), and to spatially resolve, in some cases, the ULX radio counterpart (\citealt{mezcua2013}a). However, only a limited number of ULX radio
counterparts have been found so far (\citealt{2003Sci...299..365K};
\citealt{2005A&A...436..427K}; \citealt{2005ApJ...623L.109M}; \citealt{2006A&A...452..739S}; 
\citealt{2007ApJ...666...79L}; \citealt{2011AN....332..384P}; \citealt{2012ApJ...749...17C}; \citealt{2012Sci...337..554W}; \citealt{2013Natur.493..187M}). 
Under the assumption of sub-Eddington
accretion, the radio luminosity can be used together with the X-ray
luminosity to estimate the $M_\mathrm{BH}$ by means of the radio/X-ray
correlation or the so-called `Fundamental Plane' of accreting BHs (cf. \citealt{2003A&A...400.1007C}; \citealt{2003MNRAS.344...60G};
\citealt{2003MNRAS.345.1057M}; \citealt{2004A&A...414..895F}; \citealt{2006A&A...456..439K}).
The correlation holds only for sources with steady jet emission that are in a low/hard X-ray spectral state. 
The core radio luminosity must be used in order to locate the source on the BH Fundamental Plane; hence, high-resolution radio observations are required to detect the compact radio emission.

In this paper, we present the results of a European VLBI Network (EVN\footnote{www.evlbi.org}) programme initiated with the aim of detecting
and studying milliarcsecond-scale emission in ULXs with known radio counterparts.
Four ULXs were observed with the EVN at 1.6 GHz corresponding to a wavelength $\lambda = 18$ cm. A description of this sample and the observations performed are presented in Section~\ref{observations}. The results obtained and the constraints placed on the nature of the targeted ULXs are shown in Section~\ref{results}, while the final conclusions can be found in Section~\ref{conclusions}.

\section{Sample and observations}
\label{observations}
\subsection{Sample}
\begin{figure*}
  \includegraphics[width=\textwidth]{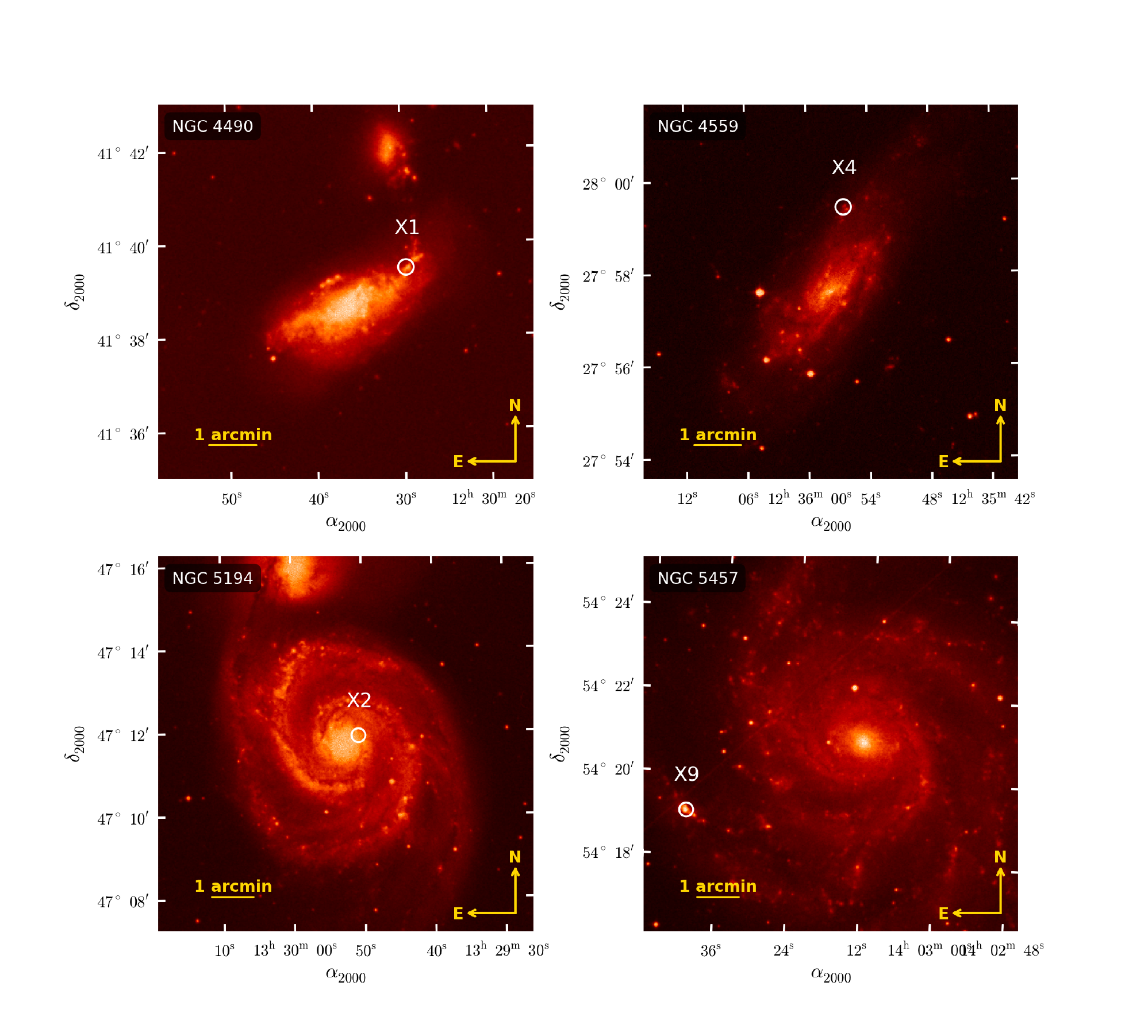}
\caption{Digitized Sky Survey (\citealt{1990AJ.....99.2019L}) images of the host galaxies of the four targeted ULXs. The ULX positions are marked with a white circle. The respective ULX names are labelled. For NGC 4490 and NGC 4559 (top--left and top--right panels, respectively), the field size is 8 arcmin $\times$ 8 arcmin. For NGC 5194 and NGC 5457 (bottom--left and bottom--right panels, respectively), the field size is 9 arcmin $\times$ 9 arcmin.}
\label{DSS}
\end{figure*}

The target sources were selected from the ULX radio catalogues of \cite{2006A&A...452..739S} and \cite{2011AN....332..384P}
as having a bright and compact Very Large Array (VLA) radio counterpart and being located in the disc of the host galaxies to avoid associations with background QSO/AGN (see Fig.~\ref{DSS}).
The identification of the ULX radio counterparts was based on a cross-correlation of the \textit{ROSAT} and \textit{Chandra} ULX catalogues (\citealt{2004ApJS..154..519S}; \citealt{2005ApJS..157...59L}) with the VLA FIRST\footnote{Faint Images of the Radio Sky at Twenty-cm} survey, which has a sensitivity of $\sim$0.15 mJy and an angular resolution of $\sim5$ arcsec (\citealt{1995ApJ...450..559B}).
A summary of the main properties of the X-ray and radio counterparts of the targets are presented in Table~\ref{table1}. The
columns listed are: ULX host galaxy name (Column 1), ULX name (Column 2), right ascension and declination at
the J2000 epoch (Columns 3 and 4, respectively) of the \textit{Chandra} counterpart, most recent \textit{Chandra} unabsorbed luminosity in the 0.3--8 keV band (fluxes estimated assuming Galactic absorption and a power-law spectrum with photon index $\Gamma = 1.7-1.9$; Col. 5), \textit{Chandra} peak luminosities in the 0.3--8 keV band (Column 6), radio (VLA FIRST) versus X-ray offset (Column 7), integrated VLA radio flux density (Column 8), type of radio emission of the VLA FIRST counterpart (Column 9), and references (Column 10).

\subsection{VLBI radio observations}
The four ULX targets were observed during two 12 h observing runs on 2012 June 2012 8 and 9 with the EVN at 1.6 GHz (experiments EM095B and EM095C). 
The participating VLBI stations for EM095C (targets N5457-X9 and N5194-X2) were Effelsberg (Germany), Westerbork (the Netherlands), Jodrell Bank (United Kingdom), Onsala (Sweden), Medicina (Italy), Noto (Italy), and Torun  (Poland). In experiment EM095B (observed targets N4490-X1 and N4559-X4) the Hartebeesthoek (South Africa) antenna also participated.

Observations were made in the phase-reference mode, interleaving observing scans between each target and a nearby, compact, strong source used as phase calibrator. 
We used J1224+4335 as the phase referencing source for N4490-X1, while J1229+4335, J1335+4542 and J1408+5613 were used for N4559-X4, N5194-X2 and N5457-X9, respectively.
A calibrator--target cycle of 4 min (1 min on the phase calibrator and 3 min on the source) was used. This resulted in a total integration time of $\sim$2 h on each target (see Table~\ref{table2}). 
Finally, we observed 4C +39.25 (a bright compact radio source) as a fringe finder and bandpass calibrator.

The data were recorded at a rate of 1024 megabit per second (Mbps), in dual-circular polarization, using eight intermediate--frequency
bands each covering 16 MHz split into 32 spectral channels. This resulted in a total bandwidth of 256 MHz. The data were later correlated at the EVN Correlator Facility at JIVE\footnote{Joint Institute for VLBI in Europe, Dwingeloo, the Netherlands.} with a correlator averaging time of 4 s. 

The correlated data were calibrated in amplitude using antenna gains and system temperatures with the AIPS\footnote{Astronomical Image Processing Software of NRAO.}, and fringe-fitted using the phase calibrators to derive delay, delay rate, and phase solutions, which were interpolated and applied to the targets. For each target the task IMAGR in AIPS was then used to make a naturally weighted image from the VLBI data.
Given the offsets between the X-ray and VLA radio positions of the targets (see Table~\ref{table1}, column~7), we performed multifield imaging in order to look for radio emission at both locations: one field was centred 
at the \textit{Chandra} position of the ULX, and another one at the position of the candidate radio counterpart determined in previous VLA observations. We imaged a field of view (FOV) of $\sim$5 arcsec for the VLA field and $\sim$1 arcsec for the \textit{Chandra} field, deriving these FOVs from the positional errors of the VLA and \textit{Chandra} ULX counterparts.
To enhance the sensitivity on angular scales $\ge$ 0.01 arcsec, the visibility data from baselines shorter than 15 $\mathrm{M}\lambda$ were used for imaging. No channel averaging was applied to the data to avoid bandwidth smearing (synthesized beam degradation away from the phase centre).
The resulting images have typical rms noise of 0.02 mJy/beam and are sensitive to emission with a brightness temperature of $\ge 1\times 10^4$ K.

\subsection{\textit{Swift} X-ray observations}
Observations with the \textit{Swift} X-ray telescope (XRT) were requested for each of our four target ULXs contemporaneous with the EVN observations. We were awarded observations of N5194-X2 (observation ID = 00037267005) and N5457-X9 (observation ID = 00032481001) on 2012 June 9 with exposure times of 4.6 and 4.3 ks, respectively. The resulting data were processed and images extracted using the online XRT data processing facility\footnote{http://www.swift.ac.uk/user\_objects/} (\citealt{2009MNRAS.397.1177E}). A faint source was detected at the \textit{Chandra} position of N5194-X2, though it is confused with a bright source coincident with the nucleus of NGC 5194. In contrast, no emission was detected at the \textit{Chandra} position of N5457-X9.

We extracted counts from a circular region with radius of 15 arcsec centred at the position of N5194-X2 and with another circular region with the same radius centred on the nuclear source in NGC 5194 to mask out emission from this bright nearby source. A background circular region with a radius of 45 arcsec was placed in a nearby source-free region, and background subtracted counts were extracted. A total of 11.2 $\pm$ 3.5 counts were extracted from the position of N5194-X2, insufficient for spectral fitting. We therefore estimated a flux and luminosity for N5194-X2 with WebPIMMS\footnote{http://heasarc.nasa.gov/Tools/w3pimms.html} using the derived count rate of 2.4 $\times$ 10$^{-3}$ counts s$^{-1}$ and an absorbed power-law model using the parameters reported in \cite{2004ApJS..154..519S}, obtaining a 0.5 - 8 keV unabsorbed luminosity of 3.8 $\times$ 10$^{39}$ erg s$^{-1}$. While we caution that this luminosity should be taken as an upper limit due to probable contamination from the nearby bright source, we note that it is not dissimilar to the value of 5.58 $\times$ 10$^{39}$ erg s$^{-1}$ measured by \cite{2004ApJS..154..519S} in the same band. Scaling to the 2 - 10 keV band using the same model, we obtain an unabsorbed luminosity of 2.5 $\times$ 10$^{39}$ erg s$^{-1}$.

For N5457-X9, we extracted counts from a circular region centred on the source position with a radius of 20$\arcsec$ and a background region of radius 45$\arcsec$ placed in a nearby source-free region. A total of 2.4 $\pm$ 1.8 counts were extracted, giving a 3$\sigma$ upper limit of 1.8 $\times$ 10$^{-3}$ counts s$^{-1}$ on the count rate. Using the column density of 3.17 $\times$ 10$^{21}$ atoms cm$^{-2}$ measured by \cite{2004MNRAS.349..404J} and a power law with $\Gamma$ = 1.7 (consistent with a low/hard--state spectrum), we estimate a 3$\sigma$ 2--10 keV upper limit of 3.9 $\times$ 10$^{38}$ erg s$^{-1}$ on the luminosity of N5457-X9 using WebPIMMS.

\section{Constraints on the ULXs}  
\label{results}
\begin{figure}
  \includegraphics[width=\columnwidth]{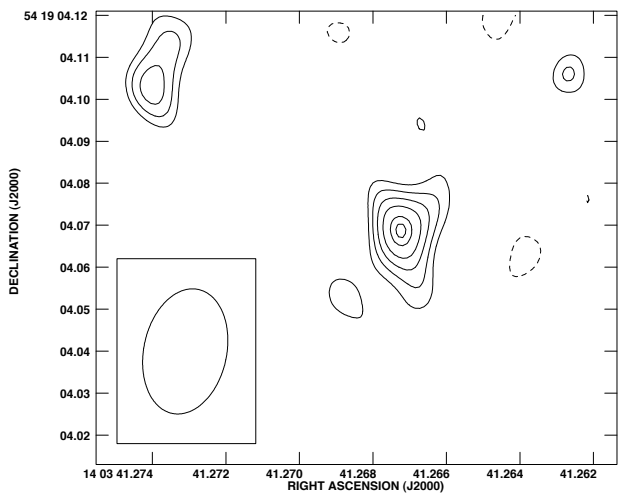}
\caption{1.6 GHz EVN image of the ULX N5457-X9 at the \textit{Chandra} position. The contours are (--3, 3, 3.5, 4, 4.5, 5 and 5.2) times the off-source rms noise of 19 $\mu$Jy beam$^{-1}$. The restoring beam size is 30 $\times$ 20 mas, with the major axis of the beam oriented along a position angle of --11$^{\circ}$.85.}
\label{fig1}
\end{figure}

For N4490-X1, N4559-X4 and N5194-X2 no radio emission was detected above a 5$\sigma$ noise level within 5 arcsec of the VLA radio source position nor within 1 arcsec of the \textit{Chandra} X-ray position. Upper limits on the flux density of these sources were obtained by estimating the local rms at the \textit{Chandra} position, and are provided in Table~\ref{table2}.
For N5457-X9, a compact component of flux density 0.12 $\pm$ 0.05 mJy\footnote{The error on the flux is not the rms noise, but indicates the combined uncertainty of fitting the emission by a Gaussian component.} is detected at a 5.2$\sigma$ level (Fig.~\ref{fig1}). This corresponds to an integrated luminosity of 1.01 $\times$ 10$^{34}$ erg s$^{-1}$ at the distance of the galaxy (6.7 Mpc; \citealt{2001ApJ...553...47F}).
The component is centred at RA(J2000) = 14$^h$03$^m$41$^s$.2672 $\pm$ 0.0003$^s$, 
Dec.(J2000) = 54$^{\circ}$19\arcmin04\arcsec.069 $\pm$ 0.003\arcsec and is coincident with the X-ray counterpart within the 0.6 arcsec \textit{Chandra} positional error. 
The fit of a two-dimensional elliptical Gaussian to the peak of emission using the AIPS task JMFIT yields a lower limit on the brightness temperature of T$_\mathrm{B}$ $>$ 8 $\times$ 10$^{4}$ K and an upper limit on the size of 32 $\times$ 22 mas.

\begin{table}
\caption{Results of the EVN observations}
\label{table2}
\begin{center}
\begin{tabular}{cccccc}
\hline
\hline 
          &                         &  \textit{Chandra} position     &      &VLA field  &    \\
\hline
Name  & Int. time & $S_\mathrm{1.6 GHz}$  &   $T_\mathrm{B}$ &    Size         &     Size    \\
          &   (h)            & (mJy beam$^{-1}$)                   &   (10$^{4}$ K)      & (arcsec)  &     (pc)      \\ 
\hline
N4490-X1	& 1.9	& $<0.020$ &  $<1.6$  &  $>0.53$   &  $>20$ \\
N4559-X4 & 1.5    & $<0.021$  &  $<1.7$  &  $>0.18$   &  $>5$ \\
N5194-X2 & 2.7   &  $<0.015$  &  $<1.2$  &  $>0.41$   &  $>17$ \\
N5457-X9 & 2.9    &  $0.12$   &  $<1.5$  &  $>0.69$   &   $>22$ \\
\hline
\end{tabular}
\end{center}
{Column designation:}~ (1) ULX name; (2) EVN integration time; (3) Peak flux density at the \textit{Chandra} position of each ULX; (4) Brightness temperature resulting from the EVN non-detections; (5, 6) Minimum sizes of the radio emitting regions in the VLA fields, in arcsec and pc respectively.
\end{table}

The non-detection of compact radio emission from N4490-X1, N4559-X4 and N5194-X2 with the EVN 
indicates that either the VLA radio emission of these ULXs is resolved out at mas scales or that the ULX radio emission is variable. The minimum sizes of the radio emitting regions in the VLA fields can be crudely estimated by considering together the VLA flux densities and the upper limits on the brightness temperature resulting from the VLBI non-detections. These limits are listed in Table~\ref{table2}.
The size of the EVN beam ($\sim30$ mas) corresponds to physical scales of $\sim1$ pc at the distances of the targeted ULXs. This rules out the presence of compact pc-scale supernova remnants (SNRs, e.g., SNR 4490-1; \citealt{mezcua2013}a) in these sources, but does not discard the presence of larger SNRs or radio nebulae resolved out by the VLBI radio observations. Indeed, the size limits inferred for these objects from the EVN data (all $>5$ pc) rather imply HII regions or very old SNRs. 
On the other hand, the presence of an accreting BH cannot be ruled out either. A decrease in radio flux at mas scales has been reported in sources with steep radio spectra, e.g., in some Seyfert nuclei (e.g., \citealt{2010MNRAS.401.2599O}), the nuclei of dwarf galaxies (e.g., \citealt{2011Natur.470...66R}) and in ULXs hosting a radio nebula (IC342-X1; e.g., \citealt{2012ApJ...749...17C}). The non-detection of mas-scale radio emission is thus still consistent with accreting BHs showing extended jet radio emission or surrounded by nebular emission, but lacking a bright compact core. 
In addition, the ULX radio emission may also be variable (e.g., \citealt{2012Sci...337..554W}; \citealt{2013Natur.493..187M}). If this is the case, the non-detections do not rule out the presence of radio jets but can be explained by the jets being temporarily turned off at the time of the observations.

Given the core radio luminosity and the X-ray luminosity of a source, an estimate of its $M_\mathrm{BH}$
can be obtained from the Fundamental Plane of accreting BHs (e.g., \citealt{2006A&A...456..439K} and references therein). This is applicable only for sources with a coupled jet-accretion disc system in the low/hard state, whose X-ray emission comes from an optically thin accretion flow or corona and is dominated by a very flat (i.e., hard) power-law spectral component (e.g., \citealt{2004A&A...414..895F}). 
Given the hard power-law X-ray spectra of three of the ULXs studied ($\Gamma$=1.7-1.9 for N4559-X4, N5194-X2 and N5457-X9) and the detection of compact radio emission in one of them (N5457-X9), the presence of a coupled jet-disc system can be assumed to estimate the $M_\mathrm{BH}$, for which we use the radio/X-ray correlation that presents the minimum scatter (i.e., the one from \citealt{2006A&A...456..439K}, with $\sigma$ = 0.12 dex; Table~\ref{table3}). 
This correlation corresponds to the upper slope of 0.63 $\pm$ 0.03 of \cite{2012MNRAS.423..590G}. It should be noted that a second track with slope 0.98 $\pm$ 0.08 has been recently found (\citealt{2012MNRAS.423..590G}), which raises the uncertainties of the BH mass estimates. 
The lack of simultaneous X-ray and radio observations can also add further sources of uncertainty; hence, simultaneous \textit{Swift} and EVN radio observations have been attempted here. We also use other data reported in the literature (i.e., the most recent \textit{Chandra} luminosities available); hence; the values presented in Table~\ref{table3}, Column~2 are meant as upper limits on the BH mass estimates. These allow us to discard SMBHs as the nature of three of the ULXs studied (N4490-X1, N4559-X4 and N5194-X2).

To further investigate the nature of the targeted ULXs, we consider each individual source in more detail, referring to published results and previous suggestions on the nature of these systems. We also derive the ratio
of 6 cm radio emission to 2--10 keV X-ray emission $R_\mathrm{X} \equiv
L_\mathrm{R}/L_\mathrm{X} = \nu L_\mathrm{\nu}(5\
\mathrm{GHz})/L_\mathrm{X}$(2-10 keV), as defined by
\cite{2003ApJ...583..145T}. We list the derived $R_\mathrm{X}$ values for each ULX in Table~\ref{table3}. We also list the $R_\mathrm{X}$ values for various objects for comparison [e.g., X-ray binaries, a
cluster of hot stars, two SNRs, and several low-luminosity (LLAGN)]. 
Using these values, we would expect IMBHs to have an $R_\mathrm{X}$ in the range --5.3 to --3.8. 

We find that the $R_\mathrm{X}$ upper limits rule out the possibility of LLAGN in N4490-X-1, N4559-X4 and N5194-X2, leaving them as viable IMBHs. For N4559-X4, the low X-ray luminosity and the upper limit on $R_\mathrm{X}$ suggest an X-ray binary nature. 
N5457-X9 is the only source for which we are able to determine a lower mass limit, which precludes the possibility of this ULX being an X-ray binary source (i.e., stellar-mass BH).

\subsection{N4490-X1}
The ULX N4490-X1 (CXOU J123029.5+413927) is one of the eight ULXs residing in the interacting galaxy pair NGC 4485/4490, at a distance of 7.8 Mpc (\citealt{1988cng..book.....T}). N4490-X1 was identified as a possible SNR due to its high absorption column, soft underlying continuum, the identification of a weakly detected [O VI] emission line and its coincidence with the radio source FIRST J123029.4+413927 (\citealt{2002MNRAS.337..677R}). However, its X-ray luminosity dropped by 30\% in a four-month period,\,a behaviour that is more typical of an X-ray binary source, and \cite{2009MNRAS.397..124G} showed that its long-term X-ray decay was inconsistent with that of a young SNR. The \textit{Chandra} X-ray spectrum was fitted by a very steep power law ($\Gamma=3.9$; \citealt{2009MNRAS.397..124G}), obtaining a 0.5-8 keV observed luminosity of 0.9 $\times$ 10$^{39}$ erg s$^{-1}$. 
This is compatible with the steep power--law/very--high state, albeit the high/soft state cannot be ruled out (e.g., \citealt{2006csxs.book..157M}). Although the Eddington fraction of this steep power-law state is not well constrained, it is known to be a sub-Eddington accretion state occurring at $\geq0.5$ Eddington rate (e.g., \citealt{2006csxs.book..157M}). Making the assumption that this source is accreting at $<100\%$ Eddington rate, we can use simple mass scaling to provide a lower mass limit of $\sim10 M_{\odot}$. If instead we assume that this source is residing in the high/soft state, which occurs at $\sim0.1$ Eddington rate, this provides a mass estimate of $\sim80 M_{\odot}$, indicating that the source is a massive stellar remnant BH.
Its VLA radio counterpart is resolved with a size of 75 pc $\times$ 150 pc and has a flat spectral index $\alpha=-0.13$ (where $F_{\nu}\propto\nu^{\alpha}$; \citealt{2004PThPS.155...27M}). The variation in both X-ray luminosity and X-ray spectral shape of N4490-X1 together with its radio emission (not detected in any of the other ULXs in the NGC 4485/4490 pair) makes it an outstanding source very different from the other ULXs in the galaxy.
To explore this difference, we will briefly assume that this source was observed in the low/hard state in order to obtain some constraints on the system from the fundamental plane of accreting BHs. To do this, we use the \textit{Chandra} luminosity taken from Swartz et al. (2011, $L_\mathrm{0.3--10.0 keV} = 9.0 \times 10^{38}$ erg s$^{-1}$), with an assumed $\Gamma$ of 1.7, scaled to the 2--10 keV bandpass using WebPIMMS. By combining this with the 5 GHz radio flux from \cite{2004PThPS.155...27M}, we find a BH mass of log $M_\mathrm{BH} < 8.3 M_{\odot}$ and a ratio log\,$R_\mathrm{X}<-2.7$.

These values are consistent with N4490-X1 being an LLAGN, but do not rule out an IMBH or X-ray binary nature.
Taking the upper limit on the EVN flux density of Table~\ref{table2} and scaling it to a 5 GHz radio luminosity using $\alpha=-0.13$, a BH mass estimate of log $M_\mathrm{BH} < 4.8 M_{\odot}$ and ratio log $R_\mathrm{X}<-4.9$ are found. These upper limits discard the presence of an SMBH, but are in agreement with N4490-X1 being either an IMBH or an X-ray binary. This result is consistent with the mass limits previously obtained from the X-ray spectral states.

\begin{table}
\caption{Estimated $M_\mathrm{BH}$ and radio/X-ray ratio of ULXs and comparison objects}
\begin{tabular}{lll}
\hline
\hline 
Name 		    & log $M_\mathrm{BH} (M_\odot)\ ^{a}$ & log $R_\mathrm{X}\ ^{b}$    \\
\hline
N4490-X1		       	&	$< 4.8$	&	$< -4.9$				\\
N4559-X4			&	$< 5.3$	&	$< -4.4$				\\
N5194-X2			&	$< 4.2$	&	$< -5.5$				\\
N5457-X9			&	$> 5.9$	&	$> -4.1$				\\
X-ray binaries	  	&      1--2		& 	$< -5.3$ $^{c}$	         	 \\
10$^{5}$ O stars at 37 Mpc & 	-- 	& 	$\sim$ 0.5 $^{d}$		  \\
SNRs	& 	-- 		&	$-2.7$ to $-2$ $^{e}$	 	   \\
LLAGN  	  		& 	6--9 		& 	$-3.8$ to $-2.8$ $^{f}$ 	    \\ 
\hline
\end{tabular}
\raggedright
\smallskip\newline
{Note:}~$^{a}$ BH mass estimated from the Fundamental Plane of \cite{2006A&A...456..439K}; $^{b}$ $R_\mathrm{X}=\mathrm\nu\,L_\mathrm{\nu}$(5\,GHz)/$L_\mathrm{X}$(2--10 keV), as defined by \cite{2003ApJ...583..145T}; $^{c}$ from \cite{2011Natur.470...66R}; $^{d}$ from \cite{2003ApJ...599.1043N}, table~2; $^{e}$ from the young SNR Cas A (\citealt{2003ApJ...599.1043N}, table~2) and J1228+441 (\citealt{2011Natur.470...66R}; \citealt{mezcua2013}a); $^{f}$ from \cite{2008ARA&A..46..475H}. 
\label{table3}
\end{table}

\subsection{N4559-X4}
N4559-X4 is located within a blue HII region close to the centre of the nearby spiral galaxy NGC 4559 (\textit{d} = 5.8 Mpc; \citealt{1992ApJS...80..479T}). It was first detected by the \textit{ROSAT} satellite (\citealt{1997A&A...318..768V}, source 8) and included in the X-ray catalogue of \cite{2005ApJS..157...59L} with a luminosity in the 0.3-8 keV band of 2.07 $\times$ 10$^{38}$ erg s$^{-1}$ (computed using a power-law spectrum with $\Gamma$ = 1.7). In later \textit{Chandra} observations (taken in 2007; \citealt{2011ApJS..192...10L}), the source had a luminosity 1.78$\times10^{38}$ erg s$^{-1}$ and its spectrum was described as being in the hard phase (\citealt{2011ApJS..192...10L}).
The source was misidentified by \cite{2006A&A...452..739S} as a ULX in the \cite{2005A&A...429.1125L} catalogue, and its low X-ray luminosity of $<10^{39}$ erg s$^{-1}$ declassifies it as a ULX.
N4559-X4 has a faint radio counterpart that does not show any extended emission
in the VLA image at 1.4 GHz (\citealt{2006A&A...452..739S}). No compact radio emission is detected with the EVN at 1.6 GHz, so an upper limit on its 1.6 GHz radio luminosity of $L_\mathrm{1.6 GHz}<1.35\times10^{33}$ erg s$^{-1}$ is derived from the rms at the position of the \textit{Chandra} counterpart. 
Despite the non-detection, the upper limit on the radio luminosity allows us to place constraints on the nature of this source. 
The VLA and EVN radio luminosities (\citealt{2006A&A...452..739S}) can be scaled to 5 GHz assuming a radio spectral index $\alpha$ = 0.15  (e.g., \citealt{2004A&A...414..895F}), while the \textit{Chandra} X-ray luminosity can be scaled to the 2-10 keV band using $\Gamma$ = 1.7 (using WebPIMMS). 
Using the compact VLA radio flux density, a BH mass of log\,$M_\mathrm{BH} < 7.8 M_{\odot}$ is derived from the location of the source in the Fundamental Plane, while the EVN upper limit further constrains the mass to log\,$M_\mathrm{BH} < 5.3 M_{\odot}$. As a result, the EVN non-detection allows us to discard an AGN as the nature of this source and suggests that N4559-X4 could host either an X-ray binary or an IMBH.
If IMBHs have values $-5.3 <$ log $R_\mathrm{X }< -3.8$, that is, in the range between those of AGN and stellar-mass BHs, the log $R_\mathrm{X }< -4.4$ derived for N4559-X4 using the EVN upper limit further discards the AGN nature and supports the X-ray binary or IMBH scenarios. 
However, the low X-ray luminosity points towards an X-ray binary as the most probable nature for this source.

\subsection{N5194-X2}
NGC\,5194 or the Whirlpool Nebula is a starburst spiral galaxy that is interacting with the galaxy NGC 5195. NGC 5194 is
located at a distance of 8.4 Mpc (\citealt{2004ApJS..154..519S}) and hosts a total of seven ULXs
(\citealt{2004ApJS..154..519S}). The ULX X2 presents a power-law X-ray spectrum of $\Gamma$ = 1.86 and luminosity $L_\mathrm{0.3--8 keV}=5.58\times10^{39}$ erg s$^{-1}$ (\citealt{2004ApJS..154..519S}), and is coincident with a peak of VLA radio emission within 0.4 arcsec
(\citealt{2011AN....332..384P}). 
From the \textit{Swift} data, an upper limit on the X-ray luminosity at the time of the EVN observation of $L_\mathrm{2-10 keV}<$ 2.5 $\times$ 10$^{39}$ erg\,s$^{-1}$ is estimated, which is the luminosity that we will use to estimate the BH mass and $R_\mathrm{X}$ ratio.
From the VLA emission, we derive an upper limit on the radio core luminosity at 5 GHz of 6.1 $\times$ 10$^{35}$ erg\,s$^{-1}$, assuming $\alpha$ = 0.15 (\citealt{2004A&A...414..895F}). Using this VLA luminosity and the \textit{Swift} luminosity, the location of N5194-X2 in the
Fundamental Plane of accreting BHs yields an estimate of the
BH mass of log\,$M_\mathrm{BH} < 7.2 M_\odot$. 
The upper limit on the compact radio emission given by the EVN observations yields a luminosity $L_\mathrm{1.6 GHz} < 2.03\times10^{33}$ erg s$^{-1}$ and allows us to further constrain the BH mass to log\,$M_\mathrm{BH} < 4.2 M_\odot$ and the $R_\mathrm{X}$ ratio to log\,$R_\mathrm{X} < -5.5$. 
The close location of the ULX to the nucleus of the host galaxy (see Fig.\ref{DSS}, bottom left) disfavours a possible background AGN, which is consistent with the EVN upper limit on the BH mass.
The log\,$R_\mathrm{X} < -5.5$ found for this ULX is consistent with an X-ray binary and lower than the $R_\mathrm{X}$ expected for an IMBH, but the uncertainties in the range of log $R_\mathrm{X}$ mean that we cannot conclusively rule out an IMBH. In the case of an X-ray binary, its X-ray luminosity ($\sim$6$\times10^{39}$ erg s$^{-1}$; see Table~\ref{table1}) could be only explained by either super-Eddington accretion, beaming effects, or a more massive stellar-mass BH (i.e. $10 < M < 100 M_\odot$).

\subsection{N5457-X9}
The ULX N5457-X9 (\citealt{2005ApJS..157...59L}; XMM-13 in \citealt{2004MNRAS.349..404J} and NGC 5457-X50 in \citealt{2011ApJS..192...10L}) is coincident with the giant HII region NGC 5461 (see Fig.~\ref{DSS}, bottom right), located in the nearby spiral galaxy M101 (or NGC 5457, \textit{d}=6.7 Mpc; \citealt{2001ApJ...553...47F}). 
This ULX was first detected by the \textit{ROSAT} satellite as a non-variable source with an X-ray luminosity in the 0.3--8.0 keV band of $3 \times 10^{38}$ erg s$^{-1}$. In later \textit{XMM--Newton} observations, the source reached an unabsorbed luminosity of $1 \times 10^{39}$ erg s$^{-1}$, and the data revealed the presence of short-term variability. The long-term lightcurve showed that the source varies over longer time-scales by a factor $\sim$5 (\citealt{2004MNRAS.349..404J,2005MNRAS.357..401J}), while its X-ray spectrum was fitted by an extremely hard power law ($\Gamma$= 0.2--0.4). However, data obtained using the \textit{Chandra} satellite (\citealt{2004MNRAS.349..404J}) were well-fitted by an absorbed cool disc blackbody model. The resultant fits from each of these models (when combined with the observed long-term lightcurve) indicate the presence of a massive stellar-mass BH or an IMBH (\citealt{2004MNRAS.349..404J}).
In the most recent (2007) \textit{Chandra} observations (\citealt{2011ApJS..192...10L}) the source presented an 
unabsorbed luminosity $L_\mathrm{0.3--8 keV}=1.23\times10^{38}$ erg s$^{-1}$ (assuming a power law of $\Gamma=1.7$ and Galactic absorption $N_\mathrm{H}=3.17 \times 10^{21}$ atoms cm$^{-2}$).
The detection of a radio counterpart to the HII region NGC 5461 with a flat spectral index of $\alpha = -0.08$ indicated that some non-thermal emission must be underlying the dominant thermal radio emission (\citealt{1990A&A...238...39G}). A cross-correlation with the VLA FIRST survey confirmed the detection of extended radio emission (\citealt{2006A&A...452..739S}). 
Here, we find that our EVN observations at 1.6 GHz reveal compact radio emission with a flux density of 0.12 mJy and brightness temperature T$_\mathrm{B}$ $>$ 8 $\times$ 10$^{4}$ K coincident with the \textit{Chandra} X-ray position within 0.6 arcsec (Fig.~\ref{fig1}). The compact source has a physical size of $\sim$1 pc, suggesting that there are two radio components present in the area around this source: an extended thermal component, as revealed by single-dish and VLA radio observations, and a compact non-thermal component.

Radio emission can be detected from HII regions, such as the HII region NGC\,5461 this source resides in, and this may be an explanation for the more extended VLA emission. 
Given the minimum size of the radio emitting region in the VLA field (see Table~\ref{table2}), the detected compact radio emission could be coming from inside an HII region of size $>$22 pc and whose extended emission is resolved out by the EVN. The compact radio detection could then come from either a compact SNR or a compact HII region inside the extended HII region NGC\,5461. Compact HII regions have been observed elsewhere, with physical sizes between 1--7 pc, their X-ray spectrum being well described by a thermal plasma model with temperatures greater than 2 keV, with a radio spectrum of spectral index $\sim$1, and brightness temperatures below 10$^{4}$ K (e.g., \citealt{2001ApJ...559..864J}; \citealt{2002MNRAS.334..912M}; \citealt{2006ApJ...653..409T}; \citealt{2007prpl.conf..181H}).
Compact SNRs also have typically physical sizes of a few pc and a steep radio spectrum dominated by optically thin synchrotron emission (e.g., \citealt{2007AJ....133.2156L}; \citealt{mezcua2013}a). However, the combination of the short- and long-term variability of N5457-X9 revealed by the X-ray data together with the X-ray power-law spectrum, the flat radio spectral index and the lower limit on the brightness temperature of 8 $\times$ 10$^{4}$ K are not consistent with a compact HII region nor an SNR\footnote{It should be noted that this flat radio spectral index was derived from single--dish observations (\citealt{1990A&A...238...39G}) and might thus be dominated by radio emission that is resolved out by the current EVN observations.}. They are more indicative of an accreting BH system. This would suggest either that we are looking at a chance coincidence of an HII region and ULX, or that the observed radio emission is from a different origin -- from the ULX itself. 

To test this, we derive the probability of a chance alignment between the VLA and the \textit{Chandra} counterparts using the number of detected sources in the FIRST survey (four sources within an FOV of 5 arcmin) and the \textit{Chandra} point spread function full width at half-maximum of 0.5 arcsec. This gives a probability of chance alignment $P$(CA) = 0.007\%. When taking only the number of sources with the same or greater flux as our counterpart (two sources), we obtain $P$(CA) = 0.004\%. Therefore, the chance alignment probability from the VLA detection of a source with the same flux or greater than our counterpart is very low. It should be noted that these $P$(CA) calculations do not take into account any potential bias introduced through our target selection or the
fact that there may be some correlation between ULXs and HII regions (as
there is a known correlation between ULXs and star formation, and also a
strong correlation between HII regions and star formation). However, while
ULX frequency appears to be correlated with SFR, only two ULXs have to date
been found to be coincident with young massive stellar clusters (see \citealt{2011MNRAS.418L.124V}; \citealt{2012ApJ...747L..13F}). It is
therefore not clear whether there really is a correlation between ULXs and
HII regions, and therefore whether or not our P(CA) values may be
underestimated.

We thus consider the possibility that both the radio and X-ray emissions are from the same source --  the ULX.
This allows us to explore the nature of N5457-X9 using the Fundamental Plane by scaling our 1.6 GHz to 5 GHz (using $\alpha = -0.08$; \citealt{1990A&A...238...39G}) and assuming that the X-ray luminosity of this source during the EVN observations was the same as that measured by \textit{Chandra} in 2007 (scaling luminosity to the 2--10 keV bandpass). This suggests that our source has a mass of log $M_\mathrm{BH} = 6.6 M_\odot$ and log $R_\mathrm{X} = -3.4$. However, the flat radio spectral index used in our scaling was derived from single-dish observations and so may be dominated by the extended emission that is resolved out in our EVN observations. Therefore, we chose to recalculate these values with $\alpha = 0.15$, a value often used to estimate BH masses from the radio/X-ray correlation (e.g., \citealt{2004A&A...414..895F}). We find that the resulting masses are similar to that derived above (log $M_\mathrm{BH} = 6.8 M_\odot$). However, these observations were not simultaneous. 

\begin{figure}
  \includegraphics[width=\columnwidth]{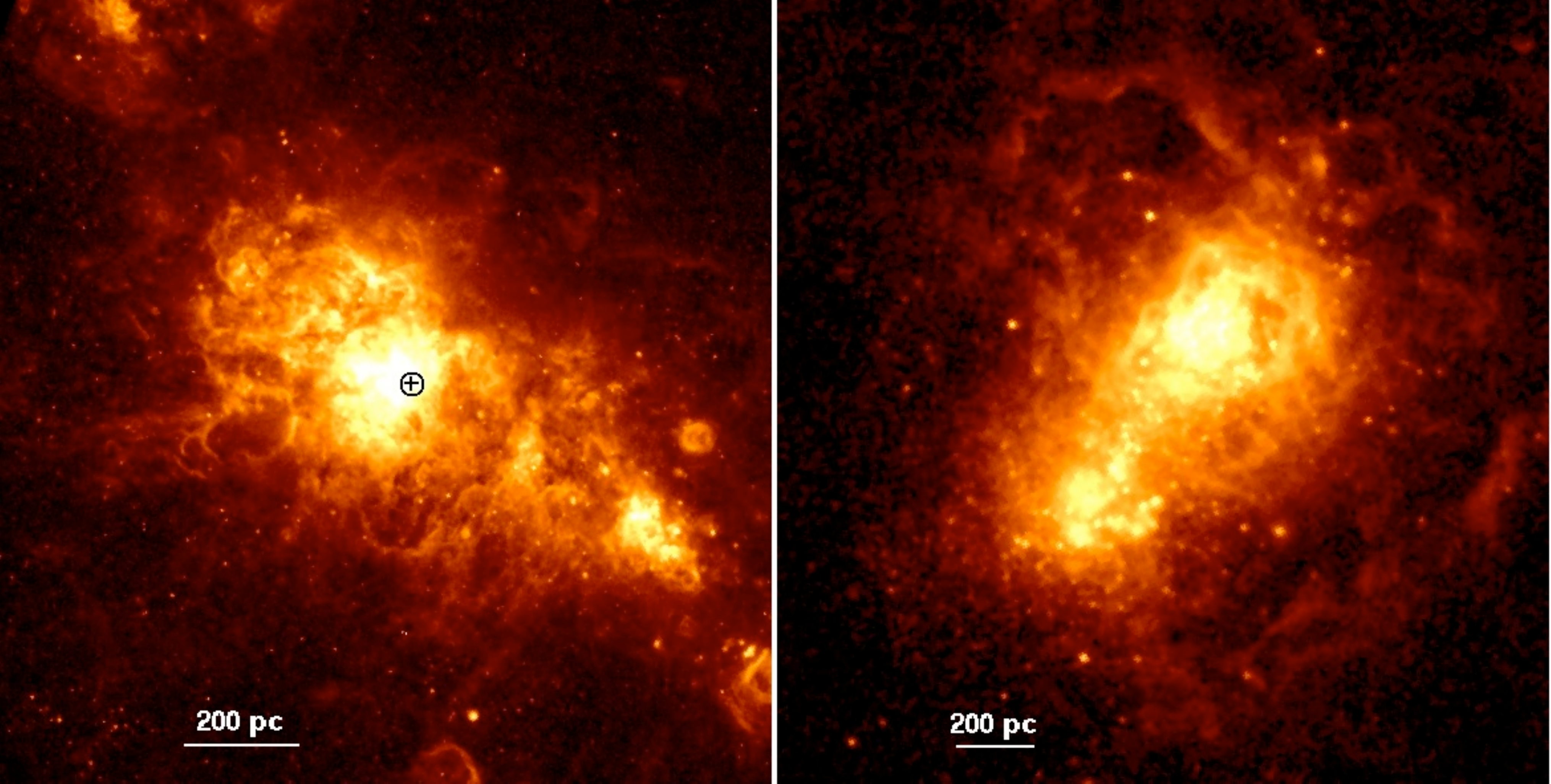}
\caption{ Left: \textit{HST F658N} image of N5457-X9, with the \textit{Chandra} position highlighted by the circle (with radius 0.6 arcsec) and the cross showing the EVN position. Right: \textit{HST F606W} image of the nearby dwarf galaxy I Zw 18.}
\label{fig2}
\end{figure}

If we now combine our radio detection with the limits obtained from our \textit{Swift} observations, we obtain more conservative limits on these values. Using the upper limit on the X-ray luminosity provided by \textit{Swift} and the EVN luminosity (scaled to 5 GHz using $\alpha = -0.08$), we estimate a lower limit on the BH mass of log $M_\mathrm{BH} > 5.9 M_\odot$ and log $R_\mathrm{X} > -4.1$, which are both consistent with either a LLAGN or a high-mass IMBH. However, the location of the ULX in a giant HII region in the outer arm of a spiral galaxy that shows no evidence for recent interactions argues very strongly against a SMBH nature. The HII region in which the ULX is located could be the remnant of a dwarf satellite galaxy that has been accreted (as has been suggested for ESO 243-49 HLX-1; \citealt{2012ApJ...747L..13F}); or it could be the IMBH formed naturally in this dense environment through the collapse of a compact massive gas cloud or successive mergers of lower mass BHs. Intriguingly, the \textit{Hubble Space Telescope} (\textit{HST}) \textit{F658N} image of this HII region shows that it has very similar filamentary emission to that seen from the dwarf galaxy I Zw 18  (see Fig.~\ref{fig2}). On the other hand, the low X-ray absorption seen in the \textit{XMM} and \textit{Chandra} data argues very strongly against this being a background object, and the lack of a bright optical point source counterpart in deep high resolution \textit{HST} images argues against it being a foreground object like a star or X-ray binary. The detection of compact radio emission consistent with the \textit{Chandra} position of N5457-X9 is thus very suggestive of either a high-mass IMBH or a low-mass SMBH, which can be explained if the giant HII region in which it is located is a satellite dwarf galaxy that is undergoing a minor merger with NGC 5457. 

\section{Conclusions}
\label{conclusions}
The association of compact radio emission with ULXs is pivotal to reveal the nature of ULXs as either SNRs, radio nebulae or accreting BHs. In this context, high-resolution (VLBI) radio observations offer a unique technique that allows us to spatially resolve the ULX radio counterpart and determine properties such as the spectral index
and brightness temperature, which gives us a hint on the size of the radio structure in case it is not resolved.
In addition, in ULXs with steady jet emission and in an X-ray low/hard state the core radio luminosity can be used together with the X-ray luminosity to estimate the ULX BH mass by means of the Fundamental Plane of accreting BHs.

With the aim of clarifying the origin of the high X-ray luminosity of four ULXs with known radio counterparts, we have observed them with the EVN at 1.6 GHz.
The results yield the detection of compact radio emission for N5457-X9, providing an accurate location of the ULX radio counterpart. The X-ray and compact radio emissions of N5457-X9 are coincident within the 0.6 arcsec \textit{Chandra} positional error, which makes N5457-X9 a strong potential high-mass IMBH candidate with steady jet radio emission. N5457-X9 thus offers a unique opportunity to test the Fundamental Plane relationship in the IMBH regime.
For the rest of the targets (N4490-X1, N4559-X4, and N5194-X2), upper limits on the radio flux density are estimated from the EVN observations. The upper limits on the BH mass estimated from the radio/X-ray correlation together with the ratio of radio to X-ray luminosities allows us to rule out the presence of an SMBH in these three sources.

\section*{Acknowledgements}
The authors are grateful to the suggestions of the anonymous referee, which helped to improve the manuscript.
The authors would like to thank A. Bahramian for his help with the X-ray data reduction and J. Callingham for his useful discussions.
S.A.F. is the recipient of an Australian Research Council Post-Doctoral Fellowship, funded by grant DP110102889. J.C.G. gratefully acknowledges funding from the Avadh Bhatia Fellowship and from an Alberta Ingenuity NewFaculty Award. We thank Neil Gehrels and the \textit{Swift} team for performing observations of N5194-X2 and N5457-X9. The European VLBI
Network is a joint facility of European, Chinese, South African,
and other radio astronomy institutes funded by their national
research councils. 
The Digitized Sky Surveys were produced at the Space Telescope Science Institute under U.S. Government grant NAG W-2166. The images of these surveys are based on photographic data obtained using the Oschin Schmidt Telescope on Palomar Mountain and the UK Schmidt Telescope. The plates were processed into the present compressed digital form with the permission of these institutions.
Based on observations made with the NASA/ESA \textit{Hubble Space Telescope}, obtained from the data archive at the Space Telescope Science Institute. STScI is operated by the Association of Universities for Research in Astronomy, Inc. under NASA contract NAS 5-26555.

\bibliographystyle{mn2e} 
\bibliography{../../SubmissionMNRAS/Files_resubmission_2/references}

\begin{thebibliography}{}

\bibitem[\protect\citeauthoryear{{Becker}, {White} \& {Helfand}}{{Becker}
  et~al.}{1995}]{1995ApJ...450..559B}
{Becker} R.~H.,  {White} R.~L.,    {Helfand} D.~J.,  1995, \apj, 450, 559

\bibitem[\protect\citeauthoryear{{Begelman}}{{Begelman}}{2002}]{2002ApJ...568L..97B}
{Begelman} M.~C.,  2002, \apjl, 568, L97

\bibitem[\protect\citeauthoryear{{Begelman}, {King} \& {Pringle}}{{Begelman}
  et~al.}{2006}]{2006MNRAS.370..399B}
{Begelman} M.~C.,  {King} A.~R.,    {Pringle} J.~E.,  2006, \mnras, 370, 399

\bibitem[\protect\citeauthoryear{{Belczynski}, {Bulik}, {Fryer}, {Ruiter},
  {Valsecchi}, {Vink} \& {Hurley}}{{Belczynski}
  et~al.}{2010}]{2010ApJ...714.1217B}
{Belczynski} K.,  {Bulik} T.,  {Fryer} C.~L.,  {Ruiter} A.,  {Valsecchi} F.,
  {Vink} J.~S.,    {Hurley} J.~R.,  2010, \apj, 714, 1217

\bibitem[\protect\citeauthoryear{{Bellovary}, {Governato}, {Quinn}, {Wadsley},
  {Shen} \& {Volonteri}}{{Bellovary} et~al.}{2010}]{2010ApJ...721L.148B}
{Bellovary} J.~M.,  {Governato} F.,  {Quinn} T.~R.,  {Wadsley} J.,  {Shen} S.,
    {Volonteri} M.,  2010, \apjl, 721, L148

\bibitem[\protect\citeauthoryear{{Colbert} \& {Mushotzky}}{{Colbert} \&
  {Mushotzky}}{1999}]{1999ApJ...519...89C}
{Colbert} E.~J.~M.,  {Mushotzky} R.~F.,  1999, \apj, 519, 89

\bibitem[\protect\citeauthoryear{{Corbel}, {Nowak}, {Fender}, {Tzioumis} \&
  {Markoff}}{{Corbel} et~al.}{2003}]{2003A&A...400.1007C}
{Corbel} S.,  {Nowak} M.~A.,  {Fender} R.~P.,  {Tzioumis} A.~K.,    {Markoff}
  S.,  2003, \aap, 400, 1007

\bibitem[\protect\citeauthoryear{{Cseh} et~al.,}{{Cseh}
  et~al.}{2012}]{2012ApJ...749...17C}
{Cseh} D.  et~al., 2012, \apj, 749, 17

\bibitem[\protect\citeauthoryear{{Cseh}, {Gris{\'e}}, {Corbel} \&
  {Kaaret}}{{Cseh} et~al.}{2011}]{2011ApJ...728L...5C}
{Cseh} D.,  {Gris{\'e}} F.,  {Corbel} S.,    {Kaaret} P.,  2011, \apjl, 728, L5

\bibitem[\protect\citeauthoryear{{Cseh}, {Lang}, {Corbel}, {Kaaret} \&
  {Gris{\'e}}}{{Cseh} et~al.}{2011}]{2011IAUS..275..325C}
{Cseh} D.,  {Lang} C.,  {Corbel} S.,  {Kaaret} P.,    {Gris{\'e}} F.,  2011, in
  {G.~E.~Romero, R.~A.~Sunyaev, \& T.~Belloni} ed.,  IAU Symposium Vol. 275,
  IAU Symposium. pp 325--326

\bibitem[\protect\citeauthoryear{{Davis}, {Narayan}, {Zhu}, {Barret},
  {Farrell}, {Godet}, {Servillat} \& {Webb}}{{Davis}
  et~al.}{2011}]{2011ApJ...734..111D}
{Davis} S.~W.,  {Narayan} R.,  {Zhu} Y.,  {Barret} D.,  {Farrell} S.~A.,
  {Godet} O.,  {Servillat} M.,    {Webb} N.~A.,  2011, \apj, 734, 111

\bibitem[\protect\citeauthoryear{{Evans} et~al.,}{{Evans}
  et~al.}{2009}]{2009MNRAS.397.1177E}
{Evans} P.~A.  et~al., 2009, \mnras, 397, 1177

\bibitem[\protect\citeauthoryear{{Fabbiano}}{{Fabbiano}}{1989}]{1989ARA&A..27...87F}
{Fabbiano} G.,  1989, \araa, 27, 87

\bibitem[\protect\citeauthoryear{{Falcke}, {K{\"o}rding} \& {Markoff}}{{Falcke}
  et~al.}{2004}]{2004A&A...414..895F}
{Falcke} H.,  {K{\"o}rding} E.,    {Markoff} S.,  2004, \aap, 414, 895

\bibitem[\protect\citeauthoryear{{Farrell} et~al.,}{{Farrell}
  et~al.}{2012}]{2012ApJ...747L..13F}
{Farrell} S.~A.  et~al., 2012, \apjl, 747, L13

\bibitem[\protect\citeauthoryear{{Farrell}, {Webb}, {Barret}, {Godet} \&
  {Rodrigues}}{{Farrell} et~al.}{2009}]{2009Natur.460...73F}
{Farrell} S.~A.,  {Webb} N.~A.,  {Barret} D.,  {Godet} O.,    {Rodrigues}
  J.~M.,  2009, \nat, 460, 73

\bibitem[\protect\citeauthoryear{{Feng} \& {Soria}}{{Feng} \&
  {Soria}}{2011}]{2011NewAR..55..166F}
{Feng} H.,  {Soria} R.,  2011, New Astronomy Reviews, 55, 166

\bibitem[\protect\citeauthoryear{{Freedman} et~al.,}{{Freedman}
  et~al.}{2001}]{2001ApJ...553...47F}
{Freedman} W.~L.  et~al., 2001, \apj, 553, 47

\bibitem[\protect\citeauthoryear{{Gallo}, {Fender} \& {Pooley}}{{Gallo}
  et~al.}{2003}]{2003MNRAS.344...60G}
{Gallo} E.,  {Fender} R.~P.,    {Pooley} G.~G.,  2003, \mnras, 344, 60

\bibitem[\protect\citeauthoryear{{Gallo}, {Miller} \& {Fender}}{{Gallo}
  et~al.}{2012}]{2012MNRAS.423..590G}
{Gallo} E.,  {Miller} B.~P.,    {Fender} R.,  2012, \mnras, 423, 590

\bibitem[\protect\citeauthoryear{{Gladstone}, {Copperwheat}, {Heinke},
  {Roberts}, {Cartwright}, {Levan} \& {Goad}}{{Gladstone}
  et~al.}{2013}]{2013ApJS..206...14G}
{Gladstone} J.~C.,  {Copperwheat} C.,  {Heinke} C.~O.,  {Roberts} T.~P.,
  {Cartwright} T.~F.,  {Levan} A.~J.,    {Goad} M.~R.,  2013, \apjs, 206, 14

\bibitem[\protect\citeauthoryear{{Gladstone} \& {Roberts}}{{Gladstone} \&
  {Roberts}}{2009}]{2009MNRAS.397..124G}
{Gladstone} J.~C.,  {Roberts} T.~P.,  2009, \mnras, 397, 124

\bibitem[\protect\citeauthoryear{{Gladstone}, {Roberts} \& {Done}}{{Gladstone}
  et~al.}{2009}]{2009MNRAS.397.1836G}
{Gladstone} J.~C.,  {Roberts} T.~P.,    {Done} C.,  2009, \mnras, 397, 1836

\bibitem[\protect\citeauthoryear{{Godet} et~al.,}{{Godet}
  et~al.}{2012}]{2012ApJ...752...34G}
{Godet} O.  et~al., 2012, \apj, 752, 34

\bibitem[\protect\citeauthoryear{{Gon{\c c}alves} \& {Soria}}{{Gon{\c c}alves}
  \& {Soria}}{2006}]{2006MNRAS.371..673G}
{Gon{\c c}alves} A.~C.,  {Soria} R.,  2006, \mnras, 371, 673

\bibitem[\protect\citeauthoryear{{Graeve}, {Klein} \& {Wielebinski}}{{Graeve}
  et~al.}{1990}]{1990A&A...238...39G}
{Graeve} R.,  {Klein} U.,    {Wielebinski} R.,  1990, \aap, 238, 39

\bibitem[\protect\citeauthoryear{{Heil}, {Vaughan} \& {Roberts}}{{Heil}
  et~al.}{2009}]{2009MNRAS.397.1061H}
{Heil} L.~M.,  {Vaughan} S.,    {Roberts} T.~P.,  2009, \mnras, 397, 1061

\bibitem[\protect\citeauthoryear{{Ho}}{{Ho}}{2008}]{2008ARA&A..46..475H}
{Ho} L.~C.,  2008, \araa, 46, 475

\bibitem[\protect\citeauthoryear{{Hoare}, {Kurtz}, {Lizano}, {Keto} \&
  {Hofner}}{{Hoare} et~al.}{2007}]{2007prpl.conf..181H}
{Hoare} M.~G.,  {Kurtz} S.~E.,  {Lizano} S.,  {Keto} E.,    {Hofner} P.,  2007,
  Protostars and Planets V, pp 181--196

\bibitem[\protect\citeauthoryear{{Jenkins}, {Roberts}, {Warwick}, {Kilgard} \&
  {Ward}}{{Jenkins} et~al.}{2004}]{2004MNRAS.349..404J}
{Jenkins} L.~P.,  {Roberts} T.~P.,  {Warwick} R.~S.,  {Kilgard} R.~E.,
  {Ward} M.~J.,  2004, \mnras, 349, 404

\bibitem[\protect\citeauthoryear{{Jenkins}, {Roberts}, {Warwick}, {Kilgard} \&
  {Ward}}{{Jenkins} et~al.}{2005}]{2005MNRAS.357..401J}
{Jenkins} L.~P.,  {Roberts} T.~P.,  {Warwick} R.~S.,  {Kilgard} R.~E.,
  {Ward} M.~J.,  2005, \mnras, 357, 401

\bibitem[\protect\citeauthoryear{{Johnson}, {Kobulnicky}, {Massey} \&
  {Conti}}{{Johnson} et~al.}{2001}]{2001ApJ...559..864J}
{Johnson} K.~E.,  {Kobulnicky} H.~A.,  {Massey} P.,    {Conti} P.~S.,  2001,
  \apj, 559, 864

\bibitem[\protect\citeauthoryear{{Kaaret} \& {Corbel}}{{Kaaret} \&
  {Corbel}}{2009}]{2009ApJ...697..950K}
{Kaaret} P.,  {Corbel} S.,  2009, \apj, 697, 950

\bibitem[\protect\citeauthoryear{{Kaaret}, {Corbel}, {Prestwich} \&
  {Zezas}}{{Kaaret} et~al.}{2003}]{2003Sci...299..365K}
{Kaaret} P.,  {Corbel} S.,  {Prestwich} A.~H.,    {Zezas} A.,  2003, Science,
  299, 365

\bibitem[\protect\citeauthoryear{{Kaaret}, {Ward} \& {Zezas}}{{Kaaret}
  et~al.}{2004}]{2004MNRAS.351L..83K}
{Kaaret} P.,  {Ward} M.~J.,    {Zezas} A.,  2004, \mnras, 351, L83

\bibitem[\protect\citeauthoryear{{King}}{{King}}{2008}]{2008MNRAS.385L.113K}
{King} A.~R.,  2008, \mnras, 385, L113

\bibitem[\protect\citeauthoryear{{King}}{{King}}{2009}]{2009MNRAS.393L..41K}
{King} A.~R.,  2009, \mnras, 393, L41

\bibitem[\protect\citeauthoryear{{King}, {Davies}, {Ward}, {Fabbiano} \&
  {Elvis}}{{King} et~al.}{2001}]{2001ApJ...552L.109K}
{King} A.~R.,  {Davies} M.~B.,  {Ward} M.~J.,  {Fabbiano} G.,    {Elvis} M.,
  2001, \apjl, 552, L109

\bibitem[\protect\citeauthoryear{{K{\"o}rding}, {Colbert} \&
  {Falcke}}{{K{\"o}rding} et~al.}{2005}]{2005A&A...436..427K}
{K{\"o}rding} E.,  {Colbert} E.,    {Falcke} H.,  2005, \aap, 436, 427

\bibitem[\protect\citeauthoryear{{K{\"o}rding}, {Falcke} \&
  {Corbel}}{{K{\"o}rding} et~al.}{2006}]{2006A&A...456..439K}
{K{\"o}rding} E.,  {Falcke} H.,    {Corbel} S.,  2006, \aap, 456, 439

\bibitem[\protect\citeauthoryear{{K{\"o}rding}, {Falcke} \&
  {Markoff}}{{K{\"o}rding} et~al.}{2002}]{2002A&A...382L..13K}
{K{\"o}rding} E.,  {Falcke} H.,    {Markoff} S.,  2002, \aap, 382, L13

\bibitem[\protect\citeauthoryear{{Lacey}, {Goss} \& {Mizouni}}{{Lacey}
  et~al.}{2007}]{2007AJ....133.2156L}
{Lacey} C.~K.,  {Goss} W.~M.,    {Mizouni} L.~K.,  2007, \aj, 133, 2156

\bibitem[\protect\citeauthoryear{{Lang}, {Kaaret}, {Corbel} \& {Mercer}}{{Lang}
  et~al.}{2007}]{2007ApJ...666...79L}
{Lang} C.~C.,  {Kaaret} P.,  {Corbel} S.,    {Mercer} A.,  2007, \apj, 666, 79

\bibitem[\protect\citeauthoryear{{Lasker}, {Sturch}, {McLean}, {Russell},
  {Jenkner} \& {Shara}}{{Lasker} et~al.}{1990}]{1990AJ.....99.2019L}
{Lasker} B.~M.,  {Sturch} C.~R.,  {McLean} B.~J.,  {Russell} J.~L.,  {Jenkner}
  H.,    {Shara} M.~M.,  1990, \aj, 99, 2019

\bibitem[\protect\citeauthoryear{{Liu}}{{Liu}}{2011}]{2011ApJS..192...10L}
{Liu} J.,  2011, \apjs, 192, 10

\bibitem[\protect\citeauthoryear{{Liu} \& {Bregman}}{{Liu} \&
  {Bregman}}{2005}]{2005ApJS..157...59L}
{Liu} J.-F.,  {Bregman} J.~N.,  2005, \apjs, 157, 59

\bibitem[\protect\citeauthoryear{{Liu} \& {Mirabel}}{{Liu} \&
  {Mirabel}}{2005}]{2005A&A...429.1125L}
{Liu} Q.~Z.,  {Mirabel} I.~F.,  2005, \aap, 429, 1125

\bibitem[\protect\citeauthoryear{{Lobanov}}{{Lobanov}}{2008}]{2008MmSAI..79.1306L}
{Lobanov} A.~P.,  2008, \memsai, 79, 1306

\bibitem[\protect\citeauthoryear{{Lodato} \& {Natarajan}}{{Lodato} \&
  {Natarajan}}{2006}]{2006MNRAS.371.1813L}
{Lodato} G.,  {Natarajan} P.,  2006, \mnras, 371, 1813

\bibitem[\protect\citeauthoryear{{McClintock} \& {Remillard}}{{McClintock} \&
  {Remillard}}{2006}]{2006csxs.book..157M}
{McClintock} J.~E.,  {Remillard} R.~A.,  2006, {Black hole binaries}.
pp 157--213

\bibitem[\protect\citeauthoryear{{McConnell} \& {Ma}}{{McConnell} \&
  {Ma}}{2013}]{2013ApJ...764..184M}
{McConnell} N.~J.,  {Ma} C.-P.,  2013, \apj, 764, 184

\bibitem[\protect\citeauthoryear{{McDonald}, {Muxlow}, {Wills}, {Pedlar} \&
  {Beswick}}{{McDonald} et~al.}{2002}]{2002MNRAS.334..912M}
{McDonald} A.~R.,  {Muxlow} T.~W.~B.,  {Wills} K.~A.,  {Pedlar} A.,
  {Beswick} R.~J.,  2002, \mnras, 334, 912

\bibitem[\protect\citeauthoryear{{Merloni}, {Heinz} \& {di Matteo}}{{Merloni}
  et~al.}{2003}]{2003MNRAS.345.1057M}
{Merloni} A.,  {Heinz} S.,    {di Matteo} T.,  2003, \mnras, 345, 1057

\bibitem[\protect\citeauthoryear{{Mezcua} \& {Lobanov}}{{Mezcua} \&
  {Lobanov}}{2011}]{2011AN....332..379M}
{Mezcua} M.,  {Lobanov} A.~P.,  2011, Astronomische Nachrichten, 332, 379

\bibitem[\protect\citeauthoryear{{Mezcua}, {Lobanov} \&
  {Mart{\'{\i}}-Vidal}}{{Mezcua} et~al.}{2013}]{mezcua2013}
{Mezcua} M.,  {Lobanov} A.~P.,    {Mart{\'{\i}}-Vidal} I.,  2013, MNRAS in
  press, arXiv: 1309.4013

\bibitem[\protect\citeauthoryear{{Middleton} et~al.,}{{Middleton}
  et~al.}{2013}]{2013Natur.493..187M}
{Middleton} M.~J.  et~al., 2013, \nat, 493, 187

\bibitem[\protect\citeauthoryear{{Miller}, {Fabbiano}, {Miller} \&
  {Fabian}}{{Miller} et~al.}{2003}]{2003ApJ...585L..37M}
{Miller} J.~M.,  {Fabbiano} G.,  {Miller} M.~C.,    {Fabian} A.~C.,  2003,
  \apjl, 585, L37

\bibitem[\protect\citeauthoryear{{Miller}, {Mushotzky} \& {Neff}}{{Miller}
  et~al.}{2005}]{2005ApJ...623L.109M}
{Miller} N.~A.,  {Mushotzky} R.~F.,    {Neff} S.~G.,  2005, \apjl, 623, L109

\bibitem[\protect\citeauthoryear{{Mirabel} \& {Rodr{\'{\i}}guez}}{{Mirabel} \&
  {Rodr{\'{\i}}guez}}{1994}]{1994Natur.371...46M}
{Mirabel} I.~F.,  {Rodr{\'{\i}}guez} L.~F.,  1994, \nat, 371, 46

\bibitem[\protect\citeauthoryear{{Moon}, {Harrison}, {Cenko} \&
  {Shariff}}{{Moon} et~al.}{2011}]{2011ApJ...731L..32M}
{Moon} D.-S.,  {Harrison} F.~A.,  {Cenko} S.~B.,    {Shariff} J.~A.,  2011,
  \apjl, 731, L32

\bibitem[\protect\citeauthoryear{{Mushotzky}}{{Mushotzky}}{2004}]{2004PThPS.155...27M}
{Mushotzky} R.,  2004, Progress of Theoretical Physics Supplement, 155, 27

\bibitem[\protect\citeauthoryear{{Neff}, {Ulvestad} \& {Campion}}{{Neff}
  et~al.}{2003}]{2003ApJ...599.1043N}
{Neff} S.~G.,  {Ulvestad} J.~S.,    {Campion} S.~D.,  2003, \apj, 599, 1043

\bibitem[\protect\citeauthoryear{{Orienti} \& {Prieto}}{{Orienti} \&
  {Prieto}}{2010}]{2010MNRAS.401.2599O}
{Orienti} M.,  {Prieto} M.~A.,  2010, \mnras, 401, 2599

\bibitem[\protect\citeauthoryear{{Pakull} \& {Gris{\'e}}}{{Pakull} \&
  {Gris{\'e}}}{2008}]{2008AIPC.1010..303P}
{Pakull} M.~W.,  {Gris{\'e}} F.,  2008, in {Bandyopadhyay} R.~M.,  {Wachter}
  S.,  {Gelino} D.,   {Gelino} C.~R.,  eds,  American Institute of Physics
  Conference Series Vol. 1010, A Population Explosion: The Nature \& Evolution
  of X-ray Binaries in Diverse Environments. pp 303--307

\bibitem[\protect\citeauthoryear{{Pakull}, {Gris{\'e}} \& {Motch}}{{Pakull}
  et~al.}{2006}]{2006IAUS..230..293P}
{Pakull} M.~W.,  {Gris{\'e}} F.,    {Motch} C.,  2006, in {E.~J.~A.~Meurs \&
  G.~Fabbiano} ed.,  IAU Symposium Vol. 230, Populations of High Energy Sources
  in Galaxies. pp 293--297

\bibitem[\protect\citeauthoryear{{Pakull} \& {Mirioni}}{{Pakull} \&
  {Mirioni}}{2003}]{2003RMxAC..15..197P}
{Pakull} M.~W.,  {Mirioni} L.,  2003, in {J.~Arthur \& W.~J.~Henney} ed.,
  Revista Mexicana de Astronomia y Astrofisica, vol. 27 Vol. 15, Revista
  Mexicana de Astronomia y Astrofisica Conference Series. pp 197--199

\bibitem[\protect\citeauthoryear{{Patruno} \& {Zampieri}}{{Patruno} \&
  {Zampieri}}{2010}]{2010MNRAS.403L..69P}
{Patruno} A.,  {Zampieri} L.,  2010, \mnras, 403, L69

\bibitem[\protect\citeauthoryear{{P{\'e}rez-Ram{\'{\i}}rez}, {Mezcua}, {Leon}
  \& {Caballero-Garc{\'{\i}}a}}{{P{\'e}rez-Ram{\'{\i}}rez}
  et~al.}{2011}]{2011AN....332..384P}
{P{\'e}rez-Ram{\'{\i}}rez} D.,  {Mezcua} M.,  {Leon} S.,
  {Caballero-Garc{\'{\i}}a} M.~D.,  2011, Astronomische Nachrichten, 332, 384

\bibitem[\protect\citeauthoryear{{Portegies Zwart}, {Baumgardt}, {Hut},
  {Makino} \& {McMillan}}{{Portegies Zwart} et~al.}{2004}]{2004Natur.428..724P}
{Portegies Zwart} S.~F.,  {Baumgardt} H.,  {Hut} P.,  {Makino} J.,
  {McMillan} S.~L.~W.,  2004, \nat, 428, 724

\bibitem[\protect\citeauthoryear{{Poutanen}, {Lipunova}, {Fabrika}, {Butkevich}
  \& {Abolmasov}}{{Poutanen} et~al.}{2007}]{2007MNRAS.377.1187P}
{Poutanen} J.,  {Lipunova} G.,  {Fabrika} S.,  {Butkevich} A.~G.,
  {Abolmasov} P.,  2007, \mnras, 377, 1187

\bibitem[\protect\citeauthoryear{{Rappaport}, {Podsiadlowski} \&
  {Pfahl}}{{Rappaport} et~al.}{2005}]{2005MNRAS.356..401R}
{Rappaport} S.~A.,  {Podsiadlowski} P.,    {Pfahl} E.,  2005, \mnras, 356, 401

\bibitem[\protect\citeauthoryear{{Reines}, {Sivakoff}, {Johnson} \&
  {Brogan}}{{Reines} et~al.}{2011}]{2011Natur.470...66R}
{Reines} A.~E.,  {Sivakoff} G.~R.,  {Johnson} K.~E.,    {Brogan} C.~L.,  2011,
  \nat, 470, 66

\bibitem[\protect\citeauthoryear{{Revnivtsev}, {Gilfanov}, {Churazov} \&
  {Sunyaev}}{{Revnivtsev} et~al.}{2002}]{2002A&A...391.1013R}
{Revnivtsev} M.,  {Gilfanov} M.,  {Churazov} E.,    {Sunyaev} R.,  2002, \aap,
  391, 1013

\bibitem[\protect\citeauthoryear{{Roberts}, {Gladstone}, {Goulding},
  {Swinbank}, {Ward}, {Goad} \& {Levan}}{{Roberts}
  et~al.}{2011}]{2011AN....332..398R}
{Roberts} T.~P.,  {Gladstone} J.~C.,  {Goulding} A.~D.,  {Swinbank} A.~M.,
  {Ward} M.~J.,  {Goad} M.~R.,    {Levan} A.~J.,  2011, Astronomische
  Nachrichten, 332, 398

\bibitem[\protect\citeauthoryear{{Roberts}, {Gladstone}, {Hetf}, {Done} \&
  {Vaughan}}{{Roberts} et~al.}{2010}]{2010AIPC.1248..123R}
{Roberts} T.~P.,  {Gladstone} J.~C.,  {Hetf} L.~M.,  {Done} C.,    {Vaughan}
  S.~A.,  2010, X-ray Astronomy 2009; Present Status, Multi-Wavelength Approach
  and Future Perspectives, 1248, 123

\bibitem[\protect\citeauthoryear{{Roberts}, {Warwick}, {Ward} \&
  {Murray}}{{Roberts} et~al.}{2002}]{2002MNRAS.337..677R}
{Roberts} T.~P.,  {Warwick} R.~S.,  {Ward} M.~J.,    {Murray} S.~S.,  2002,
  \mnras, 337, 677

\bibitem[\protect\citeauthoryear{{Russell}, {Yang}, {Gladstone}, {Wiersema} \&
  {Roberts}}{{Russell} et~al.}{2011}]{2011AN....332..371R}
{Russell} D.~M.,  {Yang} Y.-J.,  {Gladstone} J.~C.,  {Wiersema} K.,
  {Roberts} T.~P.,  2011, Astronomische Nachrichten, 332, 371

\bibitem[\protect\citeauthoryear{{S{\'a}nchez-Sutil}, {Mu{\~n}oz-Arjonilla},
  {Mart{\'{\i}}}, {Garrido}, {P{\'e}rez-Ram{\'{\i}}rez} \&
  {Luque-Escamilla}}{{S{\'a}nchez-Sutil} et~al.}{2006}]{2006A&A...452..739S}
{S{\'a}nchez-Sutil} J.~R.,  {Mu{\~n}oz-Arjonilla} A.~J.,  {Mart{\'{\i}}} J.,
  {Garrido} J.~L.,  {P{\'e}rez-Ram{\'{\i}}rez} D.,    {Luque-Escamilla} P.,
  2006, \aap, 452, 739

\bibitem[\protect\citeauthoryear{{Servillat}, {Farrell}, {Lin}, {Godet},
  {Barret} \& {Webb}}{{Servillat} et~al.}{2011}]{2011ApJ...743....6S}
{Servillat} M.,  {Farrell} S.~A.,  {Lin} D.,  {Godet} O.,  {Barret} D.,
  {Webb} N.~A.,  2011, \apj, 743, 6

\bibitem[\protect\citeauthoryear{{Shemmer}, {Brandt}, {Netzer}, {Maiolino} \&
  {Kaspi}}{{Shemmer} et~al.}{2008}]{2008ApJ...682...81S}
{Shemmer} O.,  {Brandt} W.~N.,  {Netzer} H.,  {Maiolino} R.,    {Kaspi} S.,
  2008, \apj, 682, 81

\bibitem[\protect\citeauthoryear{{Soria}, {Motch}, {Read} \& {Stevens}}{{Soria}
  et~al.}{2004}]{2004A&A...423..955S}
{Soria} R.,  {Motch} C.,  {Read} A.~M.,    {Stevens} I.~R.,  2004, \aap, 423,
  955

\bibitem[\protect\citeauthoryear{{Strateva} \& {Komossa}}{{Strateva} \&
  {Komossa}}{2009}]{2009ApJ...692..443S}
{Strateva} I.~V.,  {Komossa} S.,  2009, \apj, 692, 443

\bibitem[\protect\citeauthoryear{{Strohmayer} \& {Mushotzky}}{{Strohmayer} \&
  {Mushotzky}}{2003}]{2003ApJ...586L..61S}
{Strohmayer} T.~E.,  {Mushotzky} R.~F.,  2003, \apjl, 586, L61

\bibitem[\protect\citeauthoryear{{Strohmayer}, {Mushotzky}, {Winter}, {Soria},
  {Uttley} \& {Cropper}}{{Strohmayer} et~al.}{2007}]{2007ApJ...660..580S}
{Strohmayer} T.~E.,  {Mushotzky} R.~F.,  {Winter} L.,  {Soria} R.,  {Uttley}
  P.,    {Cropper} M.,  2007, \apj, 660, 580

\bibitem[\protect\citeauthoryear{{Swartz}, {Ghosh}, {Tennant} \& {Wu}}{{Swartz}
  et~al.}{2004}]{2004ApJS..154..519S}
{Swartz} D.~A.,  {Ghosh} K.~K.,  {Tennant} A.~F.,    {Wu} K.,  2004, \apjs,
  154, 519

\bibitem[\protect\citeauthoryear{{Swartz}, {Soria}, {Tennant} \&
  {Yukita}}{{Swartz} et~al.}{2011}]{2011ApJ...741...49S}
{Swartz} D.~A.,  {Soria} R.,  {Tennant} A.~F.,    {Yukita} M.,  2011, \apj,
  741, 49

\bibitem[\protect\citeauthoryear{{Tao}, {Feng}, {Gris{\'e}} \& {Kaaret}}{{Tao}
  et~al.}{2011}]{2011ApJ...737...81T}
{Tao} L.,  {Feng} H.,  {Gris{\'e}} F.,    {Kaaret} P.,  2011, \apj, 737, 81

\bibitem[\protect\citeauthoryear{{Terashima} \& {Wilson}}{{Terashima} \&
  {Wilson}}{2003}]{2003ApJ...583..145T}
{Terashima} Y.,  {Wilson} A.~S.,  2003, \apj, 583, 145

\bibitem[\protect\citeauthoryear{{Tsujimoto}, {Hosokawa}, {Feigelson}, {Getman}
  \& {Broos}}{{Tsujimoto} et~al.}{2006}]{2006ApJ...653..409T}
{Tsujimoto} M.,  {Hosokawa} T.,  {Feigelson} E.~D.,  {Getman} K.~V.,    {Broos}
  P.~S.,  2006, \apj, 653, 409

\bibitem[\protect\citeauthoryear{{Tully} \& {Fisher}}{{Tully} \&
  {Fisher}}{1988}]{1988cng..book.....T}
{Tully} R.~B.,  {Fisher} J.~R.,  1988, {Catalog of Nearby Galaxies}.
Cambridge University Press

\bibitem[\protect\citeauthoryear{{Tully}, {Shaya} \& {Pierce}}{{Tully}
  et~al.}{1992}]{1992ApJS...80..479T}
{Tully} R.~B.,  {Shaya} E.~J.,    {Pierce} M.~J.,  1992, \apjs, 80, 479

\bibitem[\protect\citeauthoryear{{Vogler}, {Pietsch} \& {Bertoldi}}{{Vogler}
  et~al.}{1997}]{1997A&A...318..768V}
{Vogler} A.,  {Pietsch} W.,    {Bertoldi} F.,  1997, \aap, 318, 768

\bibitem[\protect\citeauthoryear{{Volonteri}}{{Volonteri}}{2012}]{2012Sci...337..544V}
{Volonteri} M.,  2012, Science, 337, 544

\bibitem[\protect\citeauthoryear{{Volonteri}, {Lodato} \&
  {Natarajan}}{{Volonteri} et~al.}{2008}]{2008MNRAS.383.1079V}
{Volonteri} M.,  {Lodato} G.,    {Natarajan} P.,  2008, \mnras, 383, 1079

\bibitem[\protect\citeauthoryear{{Volonteri} \& {Madau}}{{Volonteri} \&
  {Madau}}{2008}]{2008ApJ...687L..57V}
{Volonteri} M.,  {Madau} P.,  2008, \apjl, 687, L57

\bibitem[\protect\citeauthoryear{{Volonteri} \& {Rees}}{{Volonteri} \&
  {Rees}}{2005}]{2005ApJ...633..624V}
{Volonteri} M.,  {Rees} M.~J.,  2005, \apj, 633, 624

\bibitem[\protect\citeauthoryear{{Voss}, {Nielsen}, {Nelemans}, {Fraser} \&
  {Smartt}}{{Voss} et~al.}{2011}]{2011MNRAS.418L.124V}
{Voss} R.,  {Nielsen} M.~T.~B.,  {Nelemans} G.,  {Fraser} M.,    {Smartt}
  S.~J.,  2011, \mnras, 418, L124

\bibitem[\protect\citeauthoryear{{Wang}}{{Wang}}{2002}]{2002MNRAS.332..764W}
{Wang} Q.~D.,  2002, \mnras, 332, 764

\bibitem[\protect\citeauthoryear{{Webb} et~al.,}{{Webb}
  et~al.}{2012}]{2012Sci...337..554W}
{Webb} N.  et~al., 2012, Science, 337, 554

\end{thebibliography}

\label{lastpage}

\end{document}